# Suppression of vortex shedding using a slit through the circular cylinder at low Reynolds number

Alok Mishra, Ashoke De*
*Department of Aerospace Engineering, Indian Institute of Technology Kanpur, Kanpur, 208016, India.*

*Corresponding Author: ashoke@iitk.ac.in

## Abstract

The present article aims to study the suppression of vortex shedding using a passive flow control technique (slit through a circular cylinder) in the laminar regime (Re=100-500). The slit width ratio S/D (slit width/diameter) on the modified cylinder plays an essential role to control the vortex shedding. The additional flow through the slit leads to the suppression of the global instability and vortex shedding, whereas a large amount of flow through the slit drastically alters the behavior of vortex shedding. The nature of vortex shedding remains periodic for all S/D, and the root mean square (rms) value of the lift coefficient decreases (in turn, vortex shedding suppression) with S/D up to Re ≤ 300. For the range Re > 300, the root mean square (rms) value of the lift coefficient decreases up to S/D < 0.15, and the flow exhibits periodic vortex shedding, while the root mean square (rms) increases beyond S/D > 0.15 due to irregular vortex shedding downstream of cylinder. The variation of the Re for the S/D=0.20 shows bifurcation points where the flow changes its behavior from symmetric to asymmetric solution at Re=232 and again becomes symmetric at Re=304. The unsteady flow analysis over the modified cylinder also indicates the suppression in the vortex shedding; however, the analysis provides the qualitative property of suppression. The reduced order modeling, i.e. Proper Orthogonal Decomposition (POD) and Dynamic Mode Decomposition (DMD), is utilized to quantify suppression and investigate the dominant vortical structure for slit through the cylinder.

## I. INTRODUCTION

Despite the simplicity in geometry, a circular cylinder exhibits very rich flow physics in fluid mechanics. The flow past a circular cylinder is essential to understand flow physics for fundamental research and engineering applications. The flow characteristic over the bluff body depends upon the non-dimensional parameter, Reynolds number (Re). At a particular Reynolds number (Re ≥ 47), the flow over the circular cylinder starts alternate periodic vortex shedding from the top and bottom surface. The vortex-induced vibration (VIV) may appear when the buff body's vortex shedding becomes any order of the structural natural frequency. VIV study is crucial to design the offshore structure, cable-stayed bridge, chimneys, buildings, and many more practical applications. The VIV produces high-frequency vibration in the bluff body structures and exerts cyclic stress loading, leading to fatigue damage of the aerodynamic structures[1-4]. Hence, suppressing the vortex-induced vibration significantly improves the life of the bluff body structures. The suppression of the VIV can be achieved with an active and passive flow control method.

The active flow control techniques require external power input, while the passive flow control techniques utilize geometric modifications. The active flow controls are very complex and need high maintenance costs compared to the passive control techniques, which are more realistic without entailing the extra energy to the system. Therefore, the present work invokes one of the passive flow control techniques. One can note that the active flow control method for the system is beyond the scope of this study. However, the interested reader can find a detailed description of the active flow control in the various published literature[5-20]. Passive flow control is extensively utilized in civil, wind, and marine engineering structures with a low maintenance cost. The geometric modification in a passive flow technique includes surface modification with roughness, dimple, helical wire, longitudinal groove, splitter plate, slit through the cylinder, segmented and wavy trailing edge, shrouds, tripping wires, and small secondary control cylinder[21-37]. In the present study, the slit through the circular cylinder is used as passive flow control techniques for suppressing the vortex shedding, in turn VIV.

The experimental studies of the slit through cylinder reveals the interesting phenomenon about the angle variation of the slit. For the range, 0° to 45° of slit angle with the incoming flow, the flow through the slit acts as a self-injecting jet downstream of the cylinder, which increases the pressure at the rear end of the cylinder. As a result, the drag pressure reduces downstream of the cylinder, and consequently, the drag coefficient decreases with the slit angle in this range. At slit angle 45°, the effect of the slit has negligible effects on the aerodynamic properties, while the drag coefficient is almost equal to the unmodified cylinder. For the 45° to 90° slit angle, the incoming jet through the slit exerts an adverse effect on the aerodynamic properties of the cylinder. The pressure downstream of the cylinder and drag coefficient shows an increasing trend in this range. For the sub-range 75° to 90°, periodic boundary layer suction and the blowing phenomenon occur for the flow over the modified cylinder. The lift coefficient and root mean square (rms) value of the lift coefficient also follow a similar pattern with the slit angle variation as the drag coefficient[38-40].

The slit width variation (the slit angle kept at 0°) on the aerodynamic properties of the cylinder is also reported in published literature. The extra amount of flow through the slit helps to weaken the vortex shedding, in turn suppressing VIV downstream of the cylinder. The self-injection of flow increases with the S/D, but the effect of the mass injection or slit width ratio on the pressure coefficient and $C_{Lrms}$ can be divided into two zones. In the first zone, the recovery of the coefficient of pressure takes place, and $C_{Lrms}$ decreases with the S/D up to a specific value. In the second zone, the extra flow through the slit downstream of the cylinder negatively affects the aerodynamic parameters. The coefficient of pressure and $C_{Lrms}$ shows the increasing trend in this zone of S/D[41]. Baek and Karniadakis[42] have also investigated the suppression of the VIV with the slit through the cylinder. The slit width ratio (S/D) varies from 0.05 to 0.30 at slit angle 0° for the Re =500 and 1000. For Re = 500, the vortex shedding shows periodic behavior up to S/D ≤ 0.16. However, the strong jet through the slit in the wake changes vortex shedding into the complicated irregular pattern beyond S/D = 0.16. The analysis provides the critical slit width at which the vortex shedding changes its behavior for both Re =500 and Re =1000.

The analysis of Baek and Karniadakis[42] paves the way for understanding the critical slit width, which leads to some pertinent queries: (i) is this critical width also the same for the flow over cylinder below Re = 500? (ii) if yes, then the pattern for the Re < 500 is similar to the flow for Re =500 (iii) how does the unsteady flow affect the suppression of vortex shedding with the slit as a passive flow control device, (iv) and finally, how does the spatial coherence of large scale structures affect the flow physics in vortex shedding suppression? Hence, the present study attempts to address the issues mentioned above by investigating the effect of the slit width on the vortex shedding through numerical simulations while covering the laminar regime flow of the cylinder from Re = 100 to 500. The critical slit width for the different Re is also examined.

The unsteady analysis over the passive flow controlled cylinder provides a qualitative description of the suppression of vortex shedding. Bukka et al.[43] studied the suppression of vortex-induced vibration using a passive flow control device (fairing and C-type connecting device). To quantify the suppression through the energy modes and identify the vortical structure, the study employs the proper orthogonal decomposition (POD) technique. Usually, the investigation of the complex interaction between wake features needs the use of the modal decomposition technique in such flows. Moreover, the POD also helps to evaluate the dominant shedding modes[44].

In the purview of the existing literature, it is understandable that the large-scale vortical structures have a significant role in the vortex shedding mixing phenomena and, in turn, suppression of the same. On the other hand, this is quite important to understand these coherent structures in suppressing vortex shedding. They carry a large portion of the energy while containing essential flow physics embedded into them. Hence, it is worthy to investigate the characteristics of these flow structures that would shed light on the understanding of the vortex shedding, in turn, flow control. The proper orthogonal decomposition (POD) is one of the classical ways that reveal the spatial coherence amongst the structure by extracting the spatial orthogonality. Kosambi[45] first proposed this and later extended it by other researchers like Lumley[46] and Sirovich[47]. Previous works of the present group extensively used this technique for different flow configurations[48-56]. Noticeably, POD, being a powerful tool to identify spatial coherence, can not capture the phase information and miss out on the dynamic information embedded inside the system. This means the POD eigen modes are unable to resolve the small perturbation leading to instability. Later, Schmid[53] introduced a technique based on the Koopman analysis called Dynamic Mode Decomposition (DMD), capable of extracting the dynamic information of the system. The primary difference between POD and DMD lies in orthogonalization in space and time, respectively, where each DMD mode is organized based on the observed frequencies. In such an arrangement, the Eigen values of DMD represent growth or decay and oscillation frequencies of each mode. Therefore, in the present study, dynamic decomposition is also performed to extract the vortex shedding dynamics to characterize the suppression of the same. This modal decomposition analysis will shed light on the dynamics of the shear layer towards the suppression of vortex shedding characteristics.

The organization of the paper is the following: Section II presents the mathematical models used for the simulations and grid independence studies. Section III reports the detailed description of the flow structures for varying physical parameters, such as the slit width ratio, followed by the conclusions in Section IV.

## II. SIMULATION DETAILS

### A. Governing equations

The study deploys the non-dimensional governing equations for viscous Newtonian incompressible flows to investigate flow over the circular cylinder. The governing equations are recast as:

Continuity equation: $\dfrac{\partial u_i}{\partial x_i} = 0$  (2.1)

Momentum equation: $\dfrac{\partial u_i}{\partial t} + u_j \dfrac{\partial u_i}{\partial x_j} = -\dfrac{\partial(p)}{\partial x_i} + \mathrm{Re}^{-1} \dfrac{\partial^2 (u_i)}{\partial x_j^2}$  (2.2)

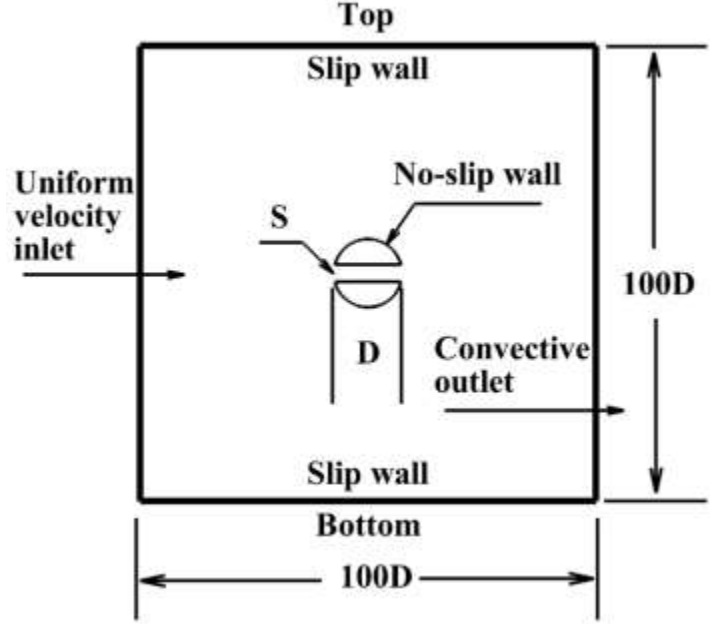

FIG. 1. Schematics of a computation domain with boundary conditions.

**B. Numerical details**

We utilize an open-source CFD code, OpenFOAM[58], to solve the fluid flow equations. The incompressible solver, pisoFoam, handles the pressure-velocity coupling of flow in the domain. We invoke Second-order discretization schemes for all the spatial and temporal terms in the governing equations while generating the two-dimensional computational grids used in the present simulations using ANSYS ICEM-CFD®. Figure 1 illustrates the square computational domain with each edge of 100D and boundary conditions for the present study. The center of the cylinder indicates the origin of the coordinate system, and the positive value of x represents the downstream of the cylinder. The inlet and outlet boundaries are placed at the $x/D = -50$ and $x/D = 50$, respectively, while the bottom and top boundaries at the $y/D = -50$ and $y = 50$, respectively.

The inlet boundary utilizes uniform flow conditions, while the outlet boundary uses the convective boundary condition. The top and bottom sides deploy slip wall conditions while the No-slip boundary condition is satisfied at both the normal and modified cylinder surface. The computational domain and boundary conditions are taken from the previously published paper by the same authors[59].

TABLE I. Grid independence test for selecting the appropriate grid for the simulations

| Grid | Cell No. | $C_D$ | % change ($C_D$) | $C_{Lrms}$ | % change ($C_{Lrms}$) |
|---|---|---|---|---|---|
| Grid-1 | 127440 | 1.3345 |  | 0.2352 |  |
| Grid-2 | 172044 | 1.3315 | 0.2248 | 0.2320 | 1.3605 |
| Grid-3 | 231000 | 1.3297 | 0.1352 | 0.2304 | 0.6896 |
| Grid-4 | 312200 | 1.3291 | 0.0451 | 0.2297 | 0.3038 |

TABLE II. Comparison of baseline cylinder case (Re=100) results with existing literature

| Available works of literature | $C_D$ | $C_{Lrms}$ | Strouhal number (St) |
|---|---|---|---|
| Park et al.[60] | 1.33 | 0.235 | 0.165 |
| Mittal[61] | 1.322 | 0.2451 | 0.164 |
| Stalberg et al.[62] | 1.32 | 0.233 | 0.166 |
| Qu et al.[63] | 1.319 | 0.225 | 0.164 |
| Posdziech and Grundmann[64] | 1.325 | 0.228 | 0.164 |
| **Present (Grid-3)** | **1.3297** | **0.2297** | **0.165** |

Table I describes the grid independence test results performed over the normal circular cylinder at Re = 100 with four successive grids. To achieve proper grid refinement over the cylinder, every consecutive grid maintains the grid ratio at 1.35. For all the different grids, the average value of $y^+$ is under one ($y^+<1$) for the resolving boundary layer on the cylinder. The mean drag coefficient (CD) results and the root means square of the lift coefficient ($C_{Lrms}$) provide the marker to investigate the sensitivity of the grid independence test. There is no significant difference in the simulated results with grid-3 and grid-4, as seen in Table I. Before selecting the final grid, we have compared the simulated results of grid-3 with the published data, as shown in Table II. The computation data provides an excellent agreement while compared with the published literature, and hence, grid-3 is considered for the detailed simulations.

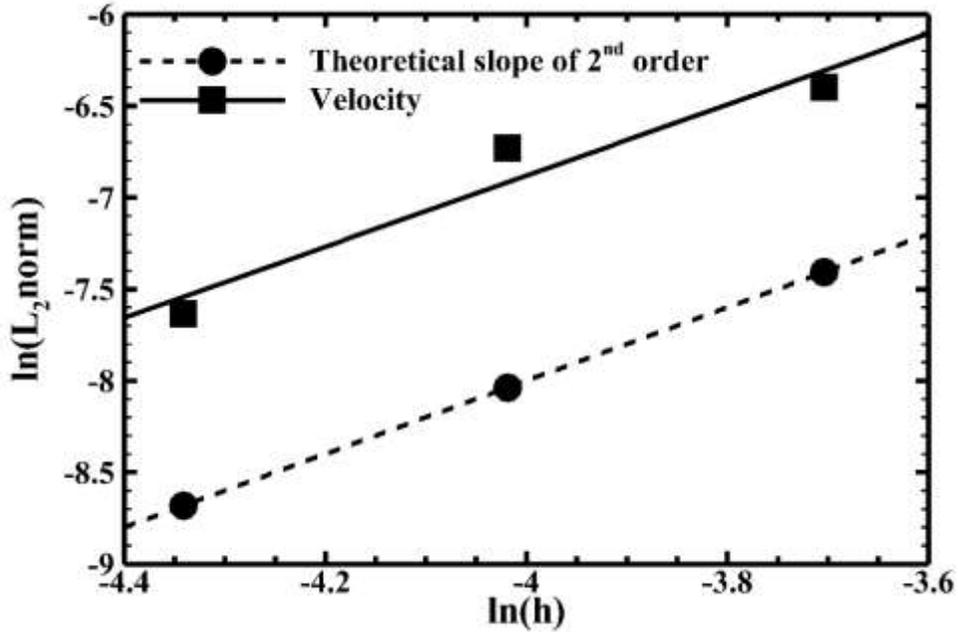

FIG. 2. Plot of $L_2$ norm against the grid spacing h

TABLE III. Richardson error estimation and Grid-Convergence Index for three sets of grids

| | $r_{32}$ | $r_{43}$ | (º) | $\varepsilon_{32}$ (10⁻²) | $\varepsilon_{43}$ (10⁻²) | R | $E^{coarse}$ (10⁻²) | $E^{fine}$ (10⁻²) | $GCI^{coarse}$ (%) | $GCI^{fine}$ (%) |
|---|---|---|---|---|---|---|---|---|---|---|
| $C_D$ | 1.35 | 1.35 | 1.94 | 0.1352 | 0.0451 | 0.3337 | -0.3063 | -0.0571 | 0.3828 | 0.0714 |
| $C_{Lrms}$ | 1.35 | 1.35 | 1.94 | 0.6896 | 0.3038 | 0.4405 | -1.5626 | -0.3846 | 1.9533 | 0.4807 |

Furthermore, the chosen grid is also cross-checked with the grid convergence index (GCI). Roache[65] proposed GCI for the uniform reporting of the grid independence studies for numerical simulations in fluids, exploring the theory of generalized Richardson extrapolation to derive the grid refinement error estimator. A detailed description of the grid convergence index (GCI) is found in Ref[59, 65]. The nomenclature of the grids is as follows: grid-2, grid-3, and grid-4 are coarse, medium, and fine grids, respectively. The convergence ratio (R) should be less than one for the monotonic convergence of the system, and it is less than one for the present work, as shown in Table III. $E$ defines the Richardson error estimator for two successive grids, and $r$ expresses the grid refinement ratio between two successive grids, while the discrete solution of two consecutive grids estimates the error ($\varepsilon$) between two grids. The $L_2$ norms of the errors between the grids provide the order of accuracy of the numerical scheme (º). Figure 2 exhibits the $L_2$ norms for the present case and the data compare with the theoretical 2nd order slope. The slope of the $L_2$ norms also refers to the order of accuracy for the numerical schemes, which is 1.94. Table III reports the GCI for the three successive grids. GCI value must be reduced for the consecutive grid increment, and in the present work, we invoke the aerodynamic parameters $C_D$ and $C_{Lrms}$ to evaluate the same. The numerical solution conclusively shows the reduction in the dependency with the refinement of the grid. GCI results also corroborate with the grid independence study (Table I and II), and we finally select grid-3 for the detailed investigation of the flow over circular cylinders.

## III. RESULTS AND DISCUSSION

We have performed 2-D simulations over the normal cylinder and modified cylinder with a slit along with the incoming flow. The simulations consist of varying Reynolds numbers, e.g. Re=100, 200, 300, 400, and 500, to obtain the insight of the slit width effect on vortex shedding in the laminar range. The slit width ratio varies from 0.05 to 0.25, with a step increment of 0.05 for a fixed slit angle at 0º.

### A. Effect of S/D on the lift coefficient

The lift coefficient shows different behavior over the Reynolds number (Re=200 and 300), as presented in Figure 3. The time history of the lift coefficient illustrates the variation against the non-dimensional time unit ($\tau = tU/D$, where t is time and U is uniform inlet velocity). The vortex shedding displays similar periodic behavior as a normal cylinder for all slit variations up to Re < 300. The slit through the cylinder acts as passive flow control, which provides an extra amount of energy to the system; thus, the pressure downstream of the cylinder increases. The increase in the pressure stabilizes the flow downstream of the cylinder, leading to a decrease in the maximum values of the lift coefficient compared to the maximum values of the lift coefficient of a normal cylinder (Figure 4). Essentially, the flow through the slit acts as a self-injecting jet that suppresses the vortex shedding.

For the range Re > 300, the variation of S/D exhibits dual behavior, which can be divided into the two zones (S/D ≤ 0.15 and S/D > 0.15). For S/D ≤ 0.15, the vortex shedding is periodic (Figure 5), and the rms (root mean square) value of lift coefficient decreases with S/D for this zone; thus, vortex shedding gets suppressed, as depicted in Figure 4. However, for S/D > 0.15, the flow shows entirely different behavior and irregular vortex shedding. The vortex interaction through the jet and main cylinder vortex occurs, which causes the loss of the periodic vortex pattern. Due to the interaction of vortices downstream of the modified cylinder, the rms value of the lift coefficient increases with the S/D, which is why the strength of vortex shedding increases. A similar trend of vortex shedding is also reported by Beck and Karniadakis[42] for Re = 500.

For Re =300, the time history of the lift coefficient suggests that the flow remains periodic against S/D variation. The pattern is different from the normal cylinder due to the interaction of the flow through the slit with the vortex generated by the cylinder at S/D = 0.20 and S/D = 0.25. The case with S/D = 0.20 shows an interesting phenomenon, where the flow exhibits asymmetric behavior downstream of the modified cylinder. A series of simulations are performed to evaluate the range of asymmetry in the flow at S/D = 0.20. Section III-D explains the detailed description of the asymmetry in the flow.

The variation of the time-averaged drag coefficient with the S/D is the same for the whole range of Reynolds number, decreasing in nature with the S/D as shown in Figure 4(b). The decrement in the drag coefficient shows steepness with the increment in the Reynolds number. For Re =500, there is a rapid reduction in the drag coefficient's value compared with other lower Reynolds number cases.

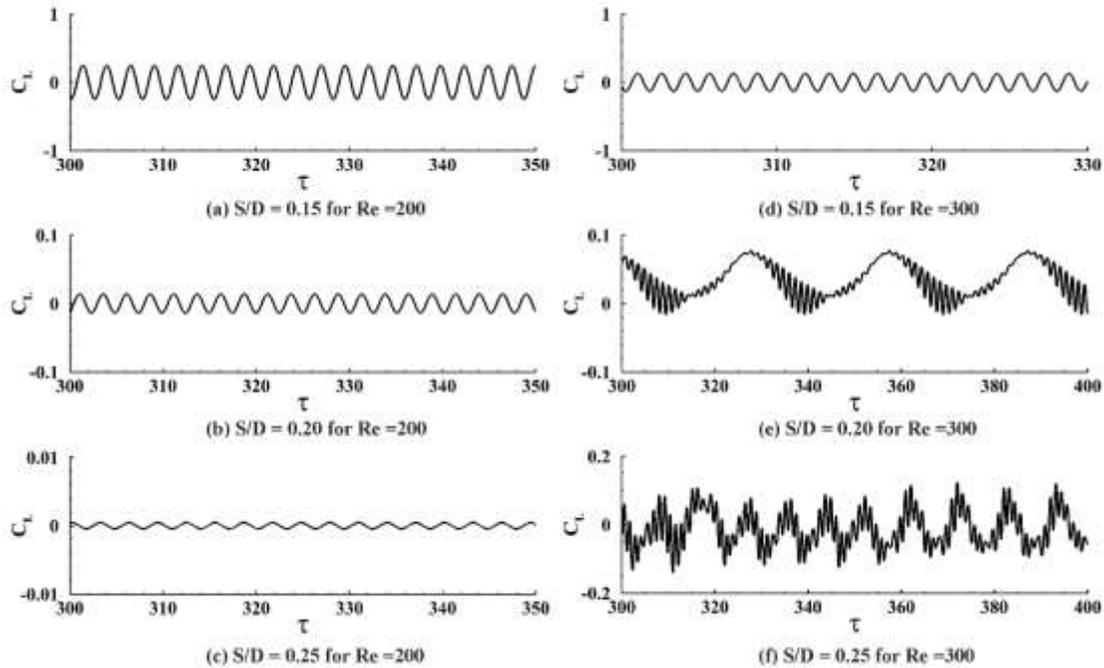

FIG. 3. The lift coefficient variation with the non-dimension time unit ($\tau$) for (a) S/D = 0.15 (b) S/D = 0.20 (c) S/D = 0.25 at Re 200, (d) S/D = 0.15 (e) S/D = 0.20 (f) S/D = 0.25 at Re = 300.

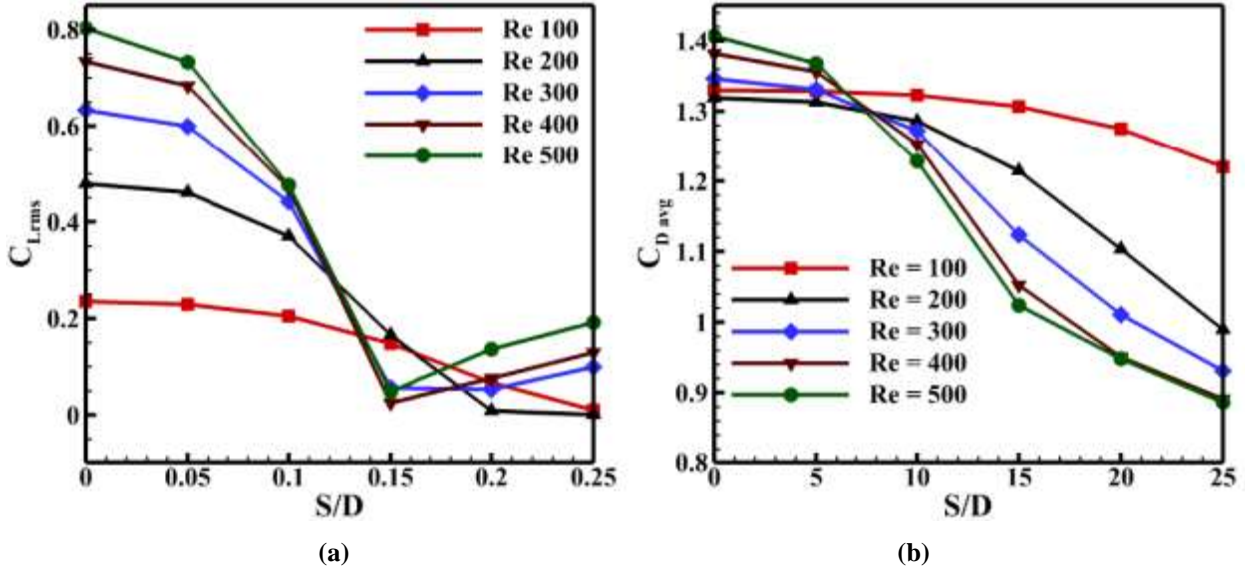

FIG. 4. (a) the rms value of lift coefficient and (b) time-averaged drag coefficient variation with S/D at various Reynolds numbers

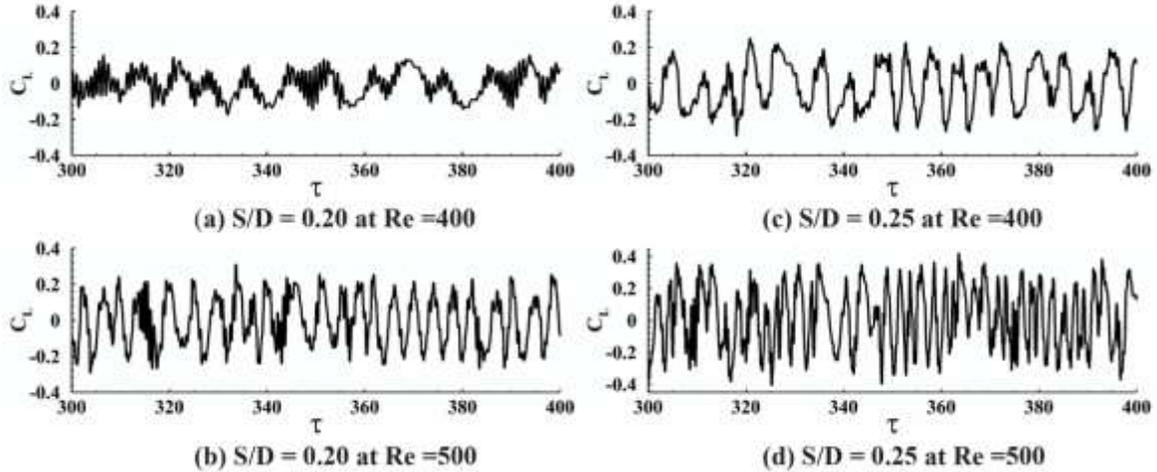

FIG. 5. Lift coefficient variation with the non-dimensional time unit ($\tau$) for (a) S/D = 0.20 at Re = 400, (b) S/D = 0.20 at Re = 500, (c) S/D = 0.25 at Re = 400 and (d) S/D =0.25 at Re = 500.

Figure 6 illustrates the fast Fourier transform of the lift coefficient (in power spectral density versus the normalized frequency) for the modified cylinder and unmodified cylinder at Re = 100 and 500. The predicted Strouhal number ($St = fD/U$, where f is frequency and U is uniform inlet velocity) for the normal cylinder at Re =100 and Re = 500 is 0.165 and 0.2118. The power spectral density for the normal cylinder contains the highest energy compared to the cases with a slit through the cylinder at Re =100, and one can note that the energy of slit cases decreases with the slit width of modified cylinders. This means the vortex shedding gets suppressed due to the presence of the slit in the cylinder, and the vortex shedding suppression is directly proportional to the slit width ratio. For Re =500, the power spectral density curve exhibits the decrement in the energy up to S/D = 0.15, and the energy of the system increases for S/D ≥ 0.15. These findings in FFT plots corroborate with the time history of the lift coefficient and the rms value of the lift coefficient.

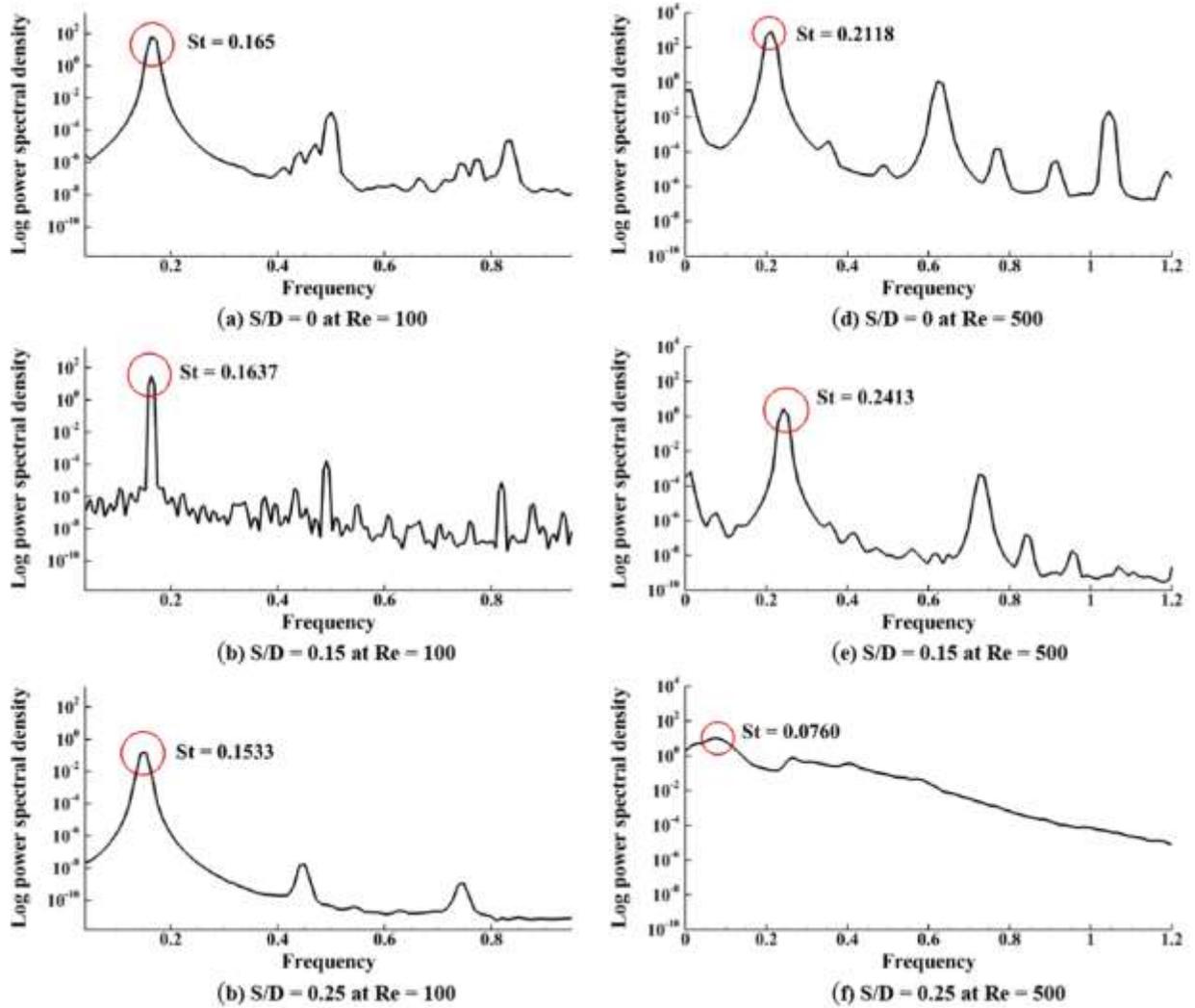

FIG. 6. Fast Fourier Transform (FFT) plots of lift coefficient values for unmodified and modified cylinder for (a) S/D = 0 (b) S/D = 0.15 (c) S/D = 0.25 at Re 100, (d) S/D = 0 (e) S/D = 0.15 (f) S/D = 0.25 at Re = 500.

### B. Effect of S/D on the global instability of flow

The slit on the cylinder acts as passive flow control, which provides an extra amount of energy to the system, reducing the global instability in flow downstream of the cylinder. The global instability of the flow can be defined in terms of the maximum value of rms fluctuation intensities over the entire computational domain[18]. The maximum values of rms fluctuation intensities of x-velocity ($u'_{rms}$) and y-velocity ($v'_{rms}$) get evaluated by extracting the values in different locations of x/D as described in Ref[59]. Figure 7 reports the maximum fluctuation intensity of $u'_{rms}$ and $v'_{rms}$ values at different Re for all S/D ratios. The maximum values of fluctuation intensities (rms value of the velocity fluctuation) reduce with an increase of the S/D, which means the global instability suppresses with the increase of the slit width ratio (S/D).

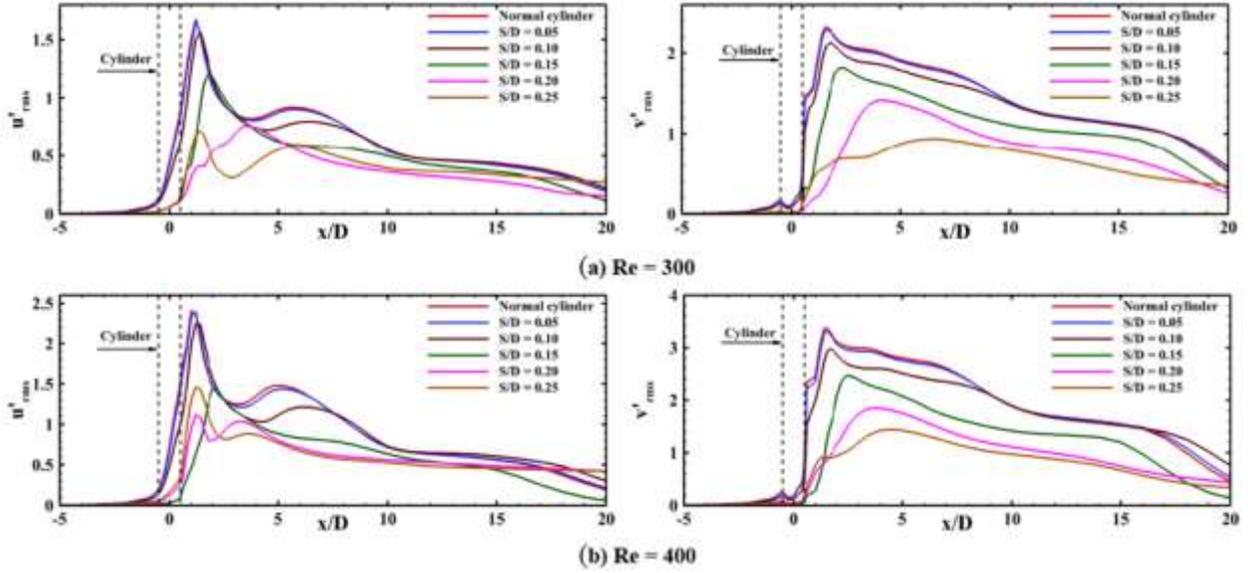

FIG. 7. The maximum value of root mean square fluctuation intensities of u′ and v′ along the line of constant x/D at (a) Re = 300 (b) Re = 400 for various S/D.

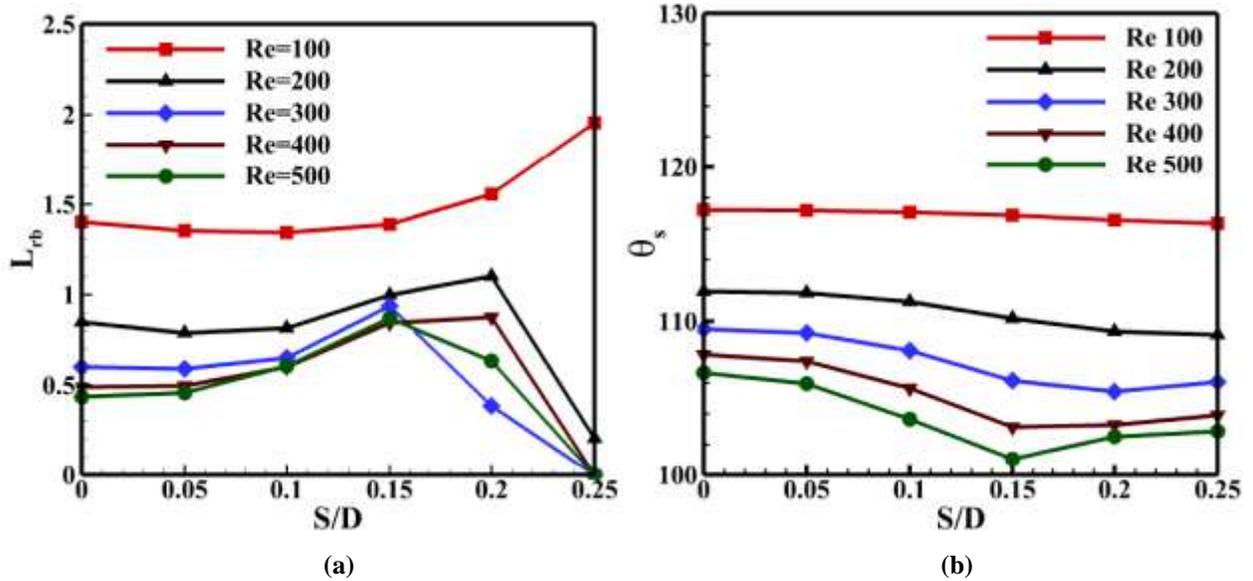

FIG. 8. (a) The normalized recirculation bubble length versus S/D for various Reynolds number and (b) Separation Angle ($\theta_s$) on the cylinder versus S/D for various Reynolds number (Measurement of angle is adopted from the author's previous publication[59])

Nishioka and Sato[66] found that the highest values point of maximum fluctuation intensity of $u'_{rms}$ and $v'_{rms}$ shift towards upstream (close to cylinder) with an increment of the Reynolds number. The maximum fluctuation intensity of $u'_{rms}$ and $v'_{rms}$ show the exponential growth inside the re-circulation bubble behind the cylinder. The zone of the exponential growth of maximum fluctuation becomes narrower at higher Reynolds numbers. Figure 7 depicts similar results for the normal cylinder as we compare the maximum fluctuation intensity growth for various Re. Figure 7 suggests that the growth of the maximum fluctuation intensity becomes gradual with the application of the slit through the cylinder. The results of the maximum fluctuation intensity of $u'_{rms}$ and $v'_{rms}$ values for the slit cylinder are equivalent to the lower Reynolds number as compared to the normal cylinder. This is the usual trend for both the maximum fluctuation intensity of $u'_{rms}$ and $v'_{rms}$ values up to Re < 400.

Surprisingly for the Re ≥ 400, the maximum fluctuation intensity of $u'_{rms}$ deviates from this trend. The growth of the maximum fluctuation intensity of $u'_{rms}$ shows gradual behavior up to S/D≤0.15, but it attains the steep slope above S/D = 0.15. One can infer that the flow over the slit cylinder for S/D ≥ 0.15 exhibits the maximum fluctuation intensity of $u'_{rms}$ values equivalent to higher Re as compared with the normal cylinder for this zone. The global instability of increases for S/D =0.25 compared to the S/D = 0.20. Grow in the instability at this range occurs due to the slit vortex and cylinder vortex interaction.

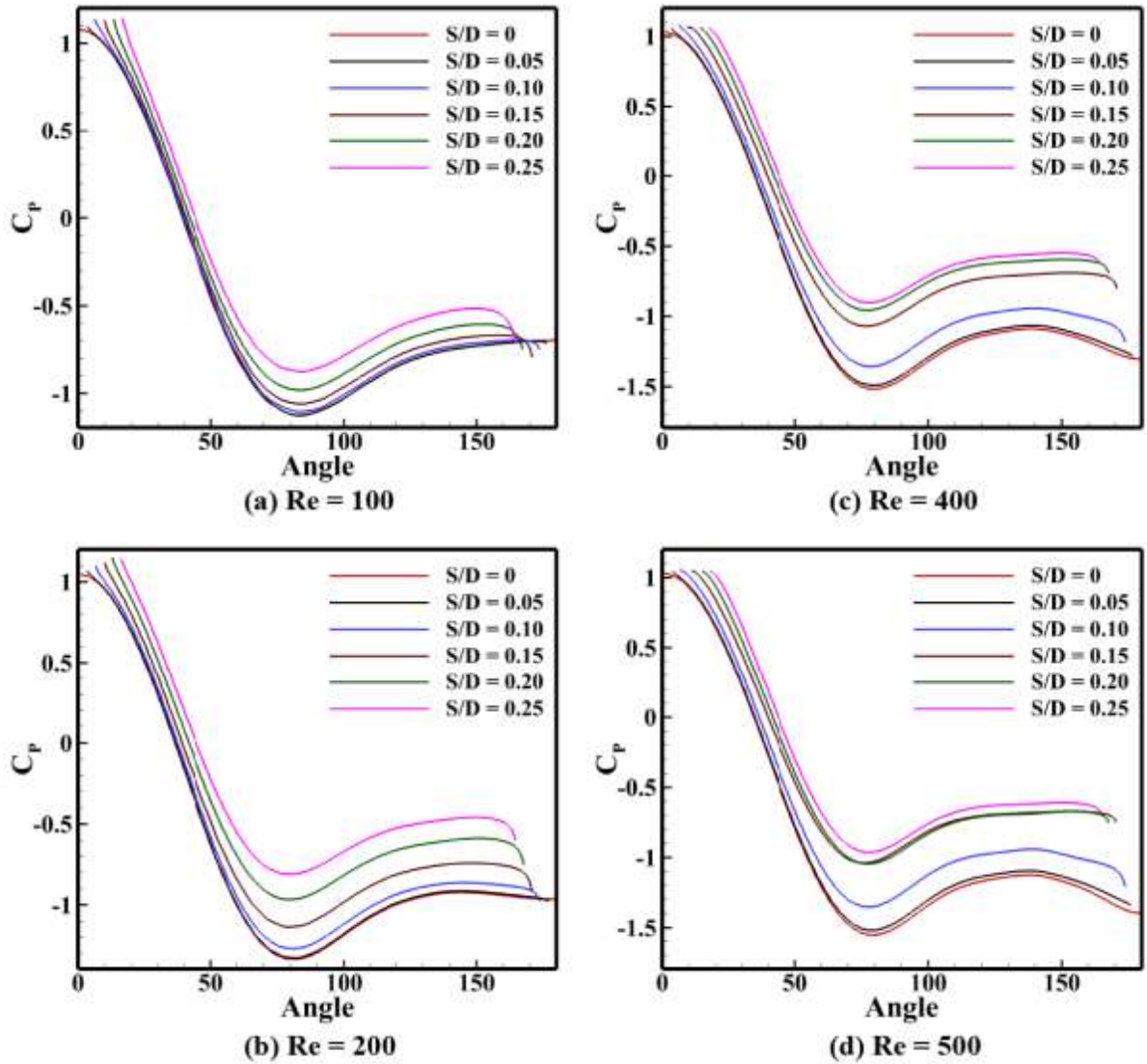

Fig. 9. Coefficient of pressure for the unmodified and modified cylinder at (a) Re = 100, (b) Re = 200, (c) Re = 400 and (d) Re = 500

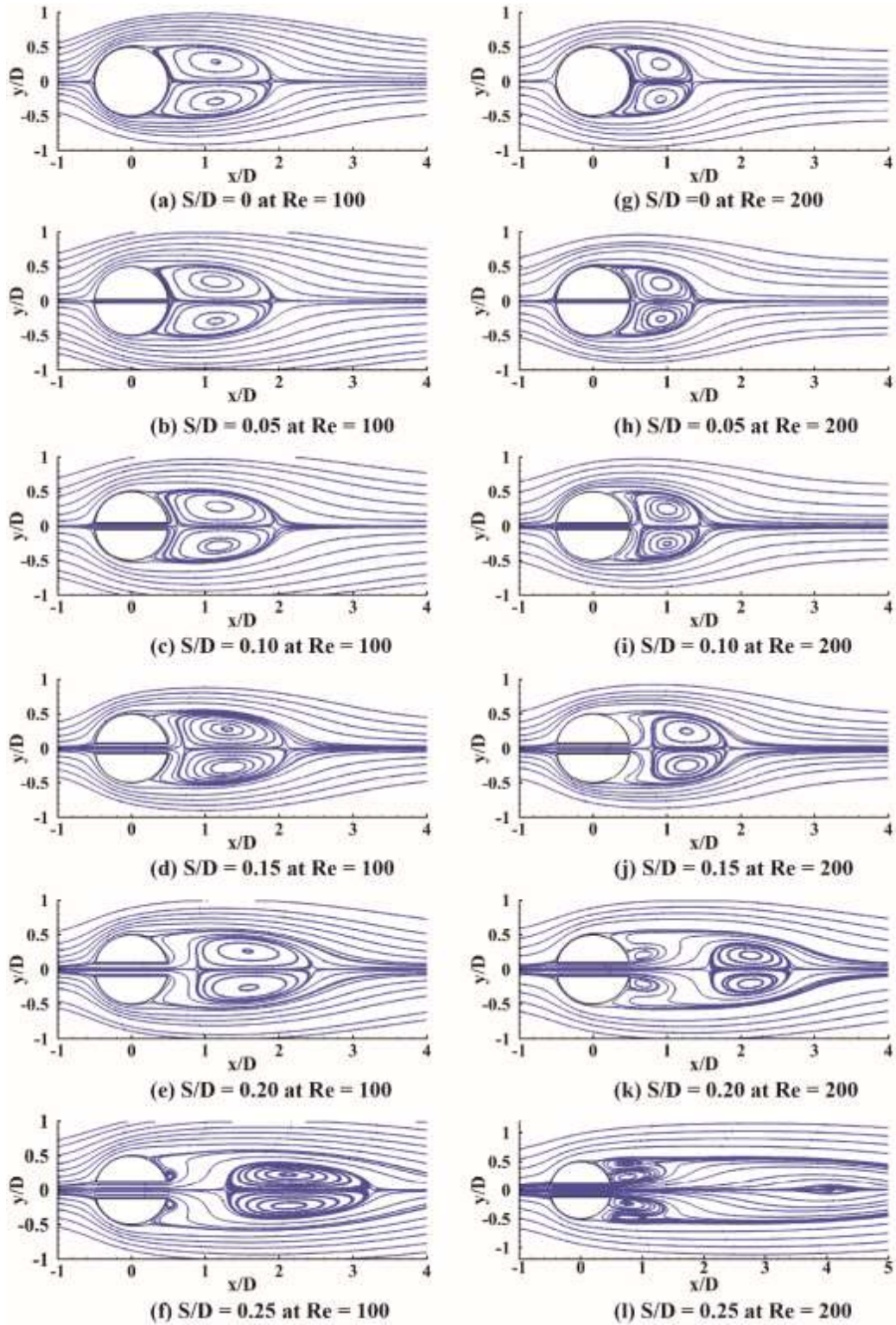

FIG. 10. Mean streamlines plots for the normal and modified cylinder for (a) S/D = 0 (b) S/D = 0.05 (c) S/D = 0.10 (d) S/D = 0.15 (e) S/D = 0.20 (f) S/D = 0.25 at Re=100 and (g) S/D = 0 (h) S/D = 0.05 (i) S/D = 0.10 (j) S/D = 0.15 (k) S/D = 0.20 (l) S/D = 0.25 at Re=200.

## C. Effect of S/D on the recirculation bubble length

Figure 8 reports the variation of the normalized recirculation bubble length ($L_{rb}$=L/D) with a slit width ratio for the range of the Reynolds number. The previous study[59] explained that the $L_{rb}$ reduces with the increment of S/D. Surprisingly, the normalized recirculation bubble length with S/D variation exhibits a different pattern from the steady-state flow range (Re < 47). The unsteady results over the normal and modified provide a distinct feature for different Reynolds numbers.

For Re =100, the $L_{rb}$ shows almost similar in size up to S/D = 0.15 and exhibits increasing trend above S/D = 0.15 (from Figure 8-a). The longest recirculation bubble length exists for Re =100 at S/D = 0.25. This point (S/D = 0.15) can be named inflection point one (IP-1) because it changes the sign of the growth of the recirculation bubble. The observation from Figures 8-b and 9-a suggest that the separation point and the pressure coefficient exhibit minor differences up to S/D=0.15, so there exists a minute change in the recirculation bubble length at this range. For S/D > 0.15, the flow through the slit influences the strength of the vortex shedding, which reduces the width of the recirculation bubble, and the bubble length gets elongated.

The separation angle also decreases, as depicted in Figure 8-b. The base suction downstream of the cylinder grows with the Re for the unmodified cylinder, which causes the instability (or maximum fluctuation intensity) increment at the recirculation region downstream of the cylinder[1,67]. The application of slit on the normal cylinder displays the decrement in the base suction (or increment of the base pressure) at the downstream of the cylinder (Figure 9-a), and the global instability also decreases with the S/D. With this logic, the flow over the slit cylinder displays the equivalent Re less than the normal cylinder's actual Re. As a result, the rms value of the lift coefficient reduces with the increase of the S/D. The lift coefficient history curve (Figure 3), the rms value of the lift coefficient (Figure 4-a), and the global instability plot (Figure 7-a) also corroborate the above discussion. Although the length of the recirculation zone increases with S/D, the width of the re-circulation zone decreases with the S/D.

The variation of the re-circulation bubble with S/D for Re =200 shows that the IP-1 exists at S/D =0.10, which can be explained through the coefficient of pressure plot (Figure 9-b). Whereas the S/D=0 to 0.10 shows almost the same base suction, but the case of S/D = 0.15 displays a sudden jump in the pressure. Furthermore, if we follow the variation of $L_{rb}$ than the curve show one more inflection point (IP-2), where the increasing trend suddenly changes sign, and decrement in the $L_{rb}$ occurs. For Re = 200, Figure 10 suggests that the two main re-circulation (or primary re-circulation) bubbles almost disappear above S/D =0.20, while S/D=0.20 happens to be the IP-2 for this case. The IP-1 and IP-2 also exist at S/D = 0.05 and S/D = 0.20 for the Re = 300-500, which resemblances with the base suction results downstream of the cylinder. A similar trend (as Re = 200) occurs for the range of Re = 300-500, and the second inflection point appears on the variation of $L_{rb}$ with S/D for Re ≥ 200 only. The case with Re=300 shows deviation from this trend due to asymmetry in the vortex shedding for the S/D = 0.20.

For Re = 400 and 500, some minor differences exist in the pressure coefficient for S/D=0.15, 0.20, and 0.25. This means the optimized vortex shedding suppression occurs at S/D=0.15, and further increment of S/D shows no significant difference in base pressure. Instead, we know from sections III-A and B that the flow displays irregular behavior, and the rms value also increases for the case S/D = 0.20 and 0.25. Figure 8-b recommends that the decreasing trend of the separation angle (with S/D) suddenly changes into an increasing trend above S/D = 0.15 for Re =400 and 500.

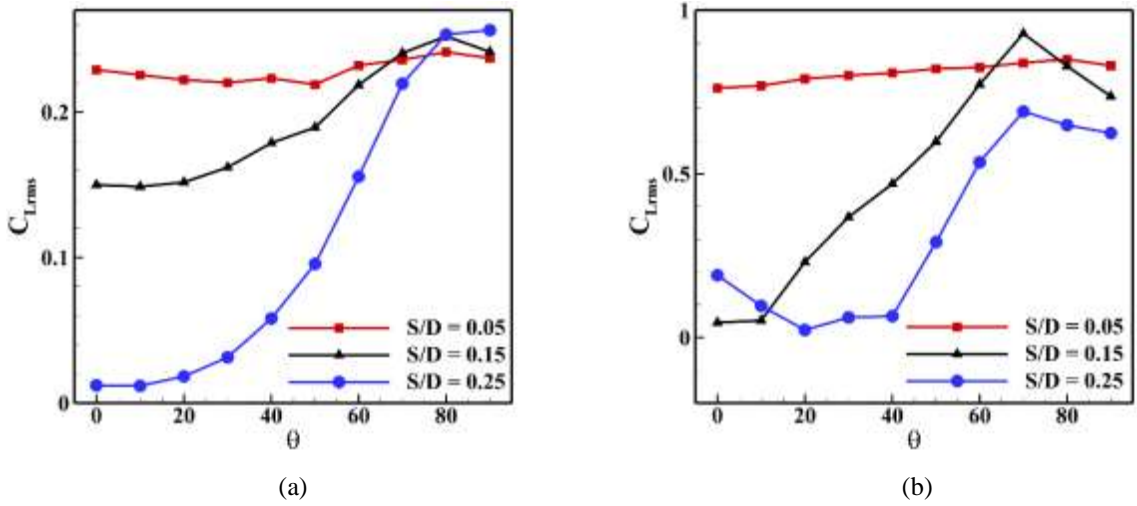

(a)                                     (b)

FIG. 11. The variation of rms value of lift coefficient with slit with ratio (a) Re = 100 and (b) Re=500.

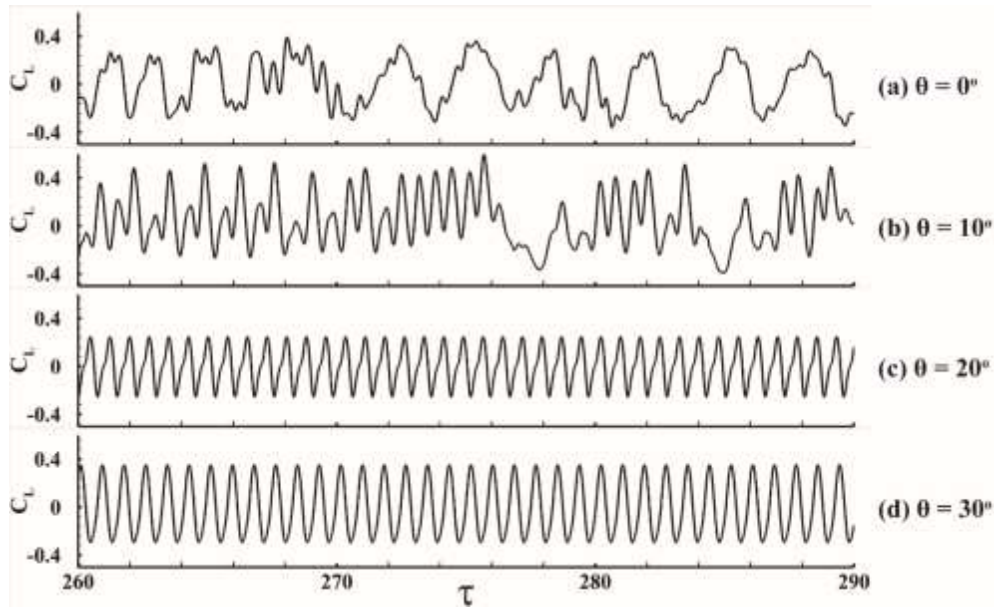

FIG. 12. Lift coefficient variation with the non-dimensional time unit ($\tau$) for various slit angles at Re=500 and S/D = 0.25.

## D. Effect of slit angle variation on the vortex shedding

To cover the whole spectrum of the slit angle variation on vortex shedding, further simulations are performed by varying the slit angle from $\theta=0°$ to $90°$ with a step angle of $\Delta\theta=10°$ for a fixed slit width ratio. As already mentioned, the fluid flow changes its property at Re =300, so two Reynolds number, i.e. Re =100 and 500 is considered to investigate the effect of slit angle variation. Figure 11 shows the variation of rms values with slit angle for various slit width ratios at Re =100 and 500. For Re =100, the flow exhibits the monotonous increment in the rms value for all slit width variations. Irrespective of the slit angle ratio variation, the flow over the slit cylinder remains periodic (not shown here). The S/D=0.25 case illustrates the maximum variation in the lift coefficient rms value.

For Re=500, the flow illustrates a similar feature as the Re=100 case (monotonously increasing with the slit angle) for the S/D=0.05 and S/D=0.15. For the S/D =0.25 case, the variation of the slit angle shows an exciting flow property over the slit cylinder. Initially, the rms value of the lift coefficient goes down with the slit angle and attains its lowest value at θ=20º. After θ=20º, the rms value of the lift coefficient starts growing up with the slit angle. Figure 12 represents the lift coefficient variation with the time for the S/D=0.25 at Re=500. As described in Section III-B, the interaction of the cylinder vortex (primary) and slit vortex (secondary) causes the increase in the global instability; hence the rms value of lift coefficient increases for S/D=0.25 as compared to S/D = 0.15. Similarly, the extra amount of flow at an angle decreases the vortex interaction (between the primary and secondary vortices), and the amount of flow also reduces through the slit cylinder. As a result, the value of the rms goes down with the slit angle up to 20°. Afterward (for θ > 20º), the flow through the slit reduces at a higher angle, causing more instability in the flow; consequently, the flow shows an increase in the rms value of lift coefficient with the slit angle up to θ=90°.

One can notice that the flow displays the complex and irregular behavior below θ=20º, and the flow becomes periodic at θ ≥ 20º (up to θ=90º – Fig. 12). The amount of extra flow downstream of the cylinder decreases as the slit angle increases. The reduction in the excess amount of flow downstream of the cylinder eventually increases the suction pressure. Subsequently, the suction pressure increase leads to increased global instability (the velocity fluctuation increases – Fig. 7). As a result, the rms value increases with the slit angle, and the extra flow through the slit goes in one direction (either lower or upper side of the cylinder). Hence, this resultant one-directional flow reduces the primary and secondary vortices interaction and attains the periodic shedding above the slit angle θ=20º.

### E. Asymmetric solution for S/D = 0.20

The variation of the time history of the lift coefficient provides an exciting phenomenon for the S/D = 0.20, as shown in Figure 13. The symmetric solution for this case exits up to Re = 230, and after that, an asymmetric solution occurs at Re = 240. One can notice that the asymmetric point (critical Reynolds number: where the flow becomes asymmetric from symmetric) exists between Re = 230-240. A series of numerical simulations are performed to evaluate the asymmetric point, and the critical Reynolds number is found to be at Re = 232 for S/D = 0.20. As described earlier, the time history of the lift coefficient at the Re = 300 provides the asymmetric in vortex shedding, and the solution at the Re = 400 shows the symmetric vortex shedding. It appears that the second critical Reynolds number lies between Re =300-400. The second critical Reynolds number is found to be at Re = 304 after a series of simulations.

The evolution of the vortex shedding with Re in Figure 13 presents another exciting phenomenon over the range of the Reynolds number. As mentioned above, the time history of the lift coefficient starts to show asymmetric flow between Re =230-240, and it can be confirmed from the lift coefficient that the flow exhibits asymmetry. For the range Re = 260-270, the vortex interaction between the slit vortex and the cylinder vortex occurs. At Re = 290, the lift coefficient shows the appearance of the second periodic vortex shedding due to the interaction of vortices, and this periodic shedding is existed up to Re =300; still, the flow remains asymmetry in this range. For the range Re = 300-320, the flow illustrates periodic vortex shedding, while the flow becomes symmetric in this range. The vortex interaction becomes strong after Re = 330, and the flow begins to show the irregular and complex shedding behavior. The source of the asymmetry in the flow for S/D= 0.20 can be explained as follows.

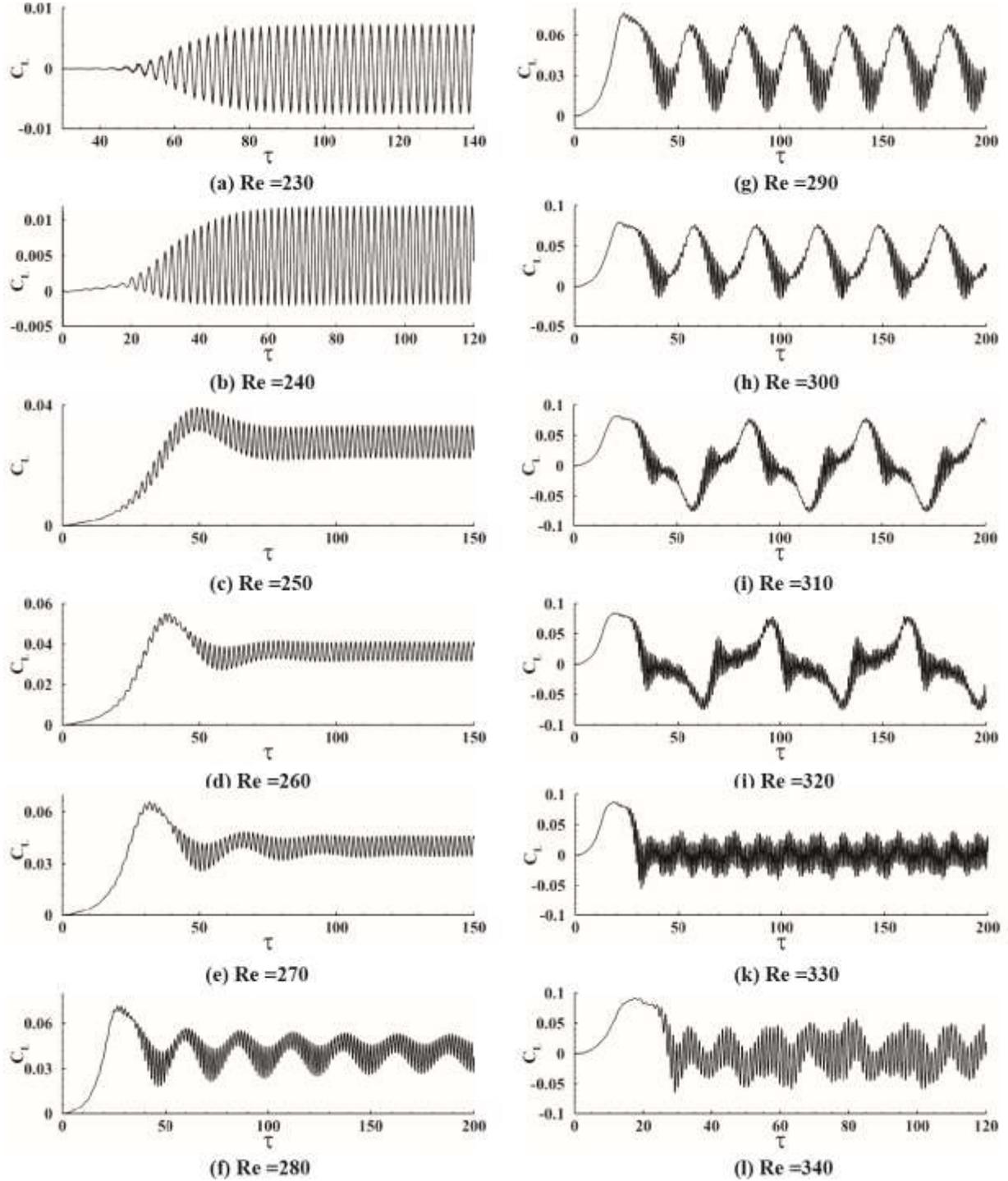

FIG. 13. Variation of the lift coefficient history at (a) Re =230, (b) Re =240, (c) Re =250, (d) Re =260, (e) Re =270, (f) Re =280, (g) Re =290, (h) Re =300, (i) Re =310, (j) Re =320, (k) Re =330 and (l) Re =340 for S/D = 0.20 case.

The flow inside the slit through the cylinder behaves like the channel flow with the Reynolds number less than the actual Reynolds number over the cylinder flow. The flow-through channel (flow through the slit) expands downstream of the cylinder. The whole system can be assumed similar to the flow-through sudden expansion channel. For sudden expansion channel with expansion ratio (ER) between 1:4 to 1:5, symmetry-breaking bifurcation occurs at Re = 35.8 (ER=1:4) and 28.5 (ER=1:5), respectively[68]. In the present scenario, for S/D = 0.20, the expansion ratio happens to be between 1:4 to 1:5 as the share layers from the top and the bottom surface of the cylinder reduce the expansion length less than the diameter of the cylinder. Correlating the same with sudden expansion channel situation, the symmetry-breaking bifurcation for S/D = 0.20 must lie between the critical Reynolds numbers (at which symmetry-breaking bifurcation takes place) for the expansion ratio 1:4 (Re = 35.8 ) and 1:5 (Re = 28.5). The effective Reynolds number inside the slit channel is Re = $U.h/\nu$, where U is the average velocity on the channel, h is the height of the channel, and $\nu$ is the kinematic viscosity. The critical Reynolds number of the slit channel (at which the bifurcation occurs) happens to be 32.60, equivalent to Re = 232 for flow over a cylinder. Evidently, the flow over the slit cylinder with S/D = 0.20 shows a similar phenomenon as the expansion channel with a ratio between 1:4 to 1:5. Moreover, the flow again displays the symmetric solution at slit channel Re = 76.42 (or Re = 304 for the flow over cylinder). The effective expansion ratio may decrease with Re, and as the effective expansion ratio decreases, the symmetry-breaking bifurcation point (Reynolds number) increases. This is the reason behind attaining symmetricity again by the flow over the slit cylinder. The asymmetry in the flow over the slit cylinder may also occur for other slit width ratios. We investigate flow symmetricity for the case of S/D =0.25 at Re = 250 (equivalent to the channel Re = 55.20).

## F. Unsteady analysis of the flow

The instantaneous vorticity around the cylinder allows us to investigate and compare the unsteady flow structure (Figures 14). The flow over the normal cylinder shows the periodic shedding, and the applying slit on the cylinder with various S/D does not change the behavior of the shedding, whereas the strength of vortices reduces with the S/D variation for Re =100. The delay in the interaction of the shear layer from both sides (upper and lower) occurs downstream of the modified cylinder compared to a normal cylinder. The extra amount of flow is responsible for the delayed interaction of the shear layers and the suppression of the vortex shedding for the modified cylinder.

One can infer from Figure 14(b) that the vortex shedding of the modified cylinder exhibits the periodic behavior for Re =500, similar to the normal cylinder up to S/D=0.15, and the maximum suppression in the vortex shedding occurs at S/D =0.15. Due to strong flow injection through the slit and the secondary vortex (through the slit of the cylinder) interact with the primary vortex, the shedding behavior of flow becomes irregular and complex. The results corroborate with the finding of section III-a.

The above analysis provides the understanding of unsteady data for the unmodified and modified cylinder only for one time instant. The study explains the suppression of vortex shedding by applying the slit through the cylinder, and one can notice the irregular and complex vortex shedding for S/D > 0.15 at Re = 500. However, the study lacks to provide details about the interaction between primary and secondary vortices in the flow over the slit cylinder. The flow over the cylinder at various time instant is considered here to explore the interaction of the vortices, and three cases (S/D = 0, 0.15, and 0.25 at Re =500) are selected for further study. Figure 15 reports the vorticity contour plots for various time instants for the normal cylinder and S/D = 0.15 at Re = 500. The difference in the secondary vortex is visual from the starting of the flow. The vortex shedding starts from time $\tau = 34$ for normal cylinder while flow shows vortex shedding from $\tau = 38$ for S/D = 0.15. The strength of vortex shedding reduces due to the interaction of primary and secondary vortices for the modified cylinder.

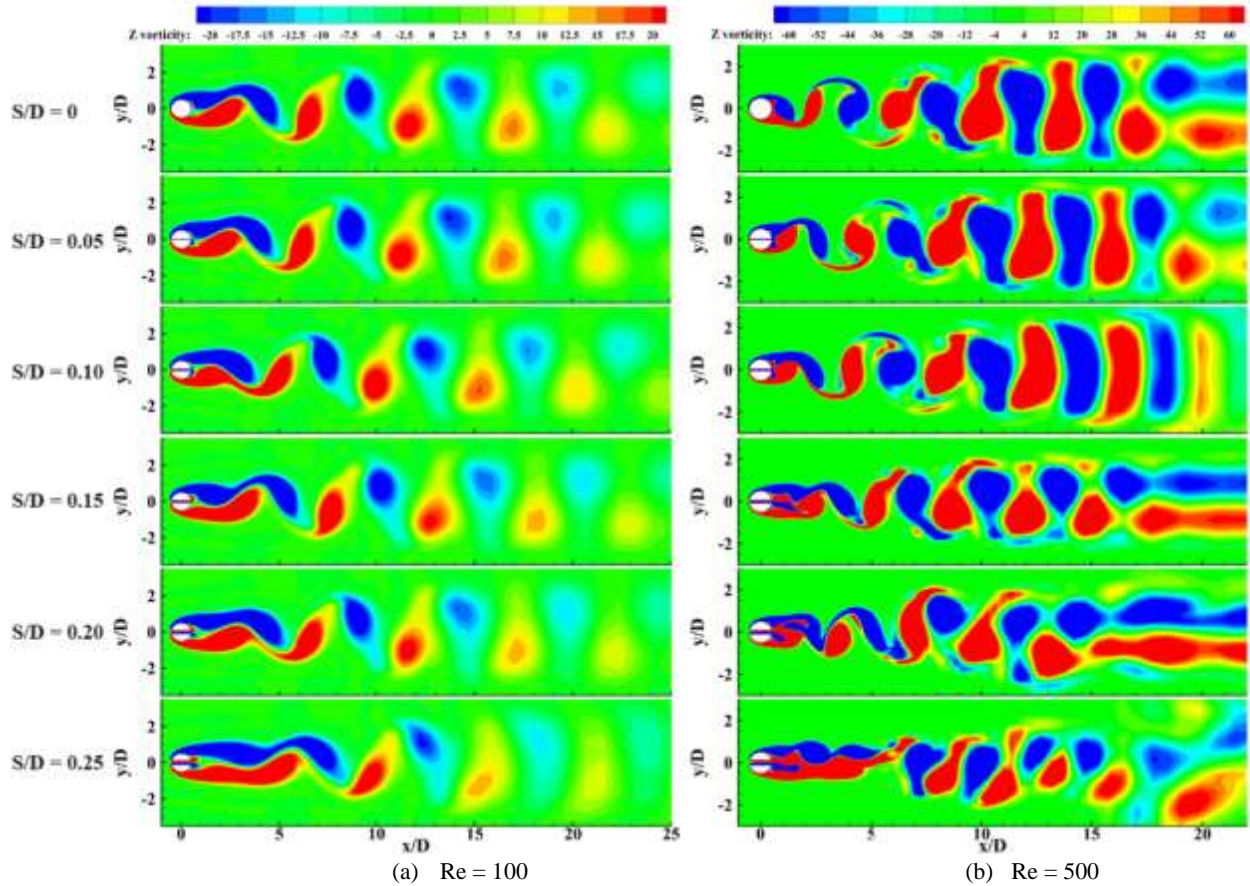

(a) Re = 100    (b) Re = 500

FIG. 14. Instantaneous vorticity contour over the normal and modified circular cylinder for for (a) S/D = 0 (b) S/D = 0.05 (c) S/D = 0.10 (d) S/D = 0.15 (e) S/D = 0.20 (f) S/D = 0.25 at Re=100 and (g) S/D = 0 (h) S/D = 0.05 (i) S/D = 0.10 (j) S/D = 0.15 (k) S/D = 0.20 (l) S/D = 0.25 at Re=500.

Figure 16 reveals a very interesting flow features for S/D = 0.25 at Re =500. The flow demonstrates two vortex shedding, one from the slit through the cylinder and the other from the cylinder surface. Initially, flow shows symmetry along with the centerline of the cylinder. The interaction of the vortices takes place after 15 flow through time. The vortex interaction of primary and secondary vortices causes the irregular and complex nature of the vortex shedding. For these instantaneous analyses, one can see the vortical structures in the presence of slit are primarily responsible for suppressing the vortex shedding downstream. However, this analysis is insufficient to detect the vortical structures and the energy mode associated with slit-vortex interactions. Hence, we expand our analysis (following subsections) using the modal decomposition of the flow field to quantify the same.

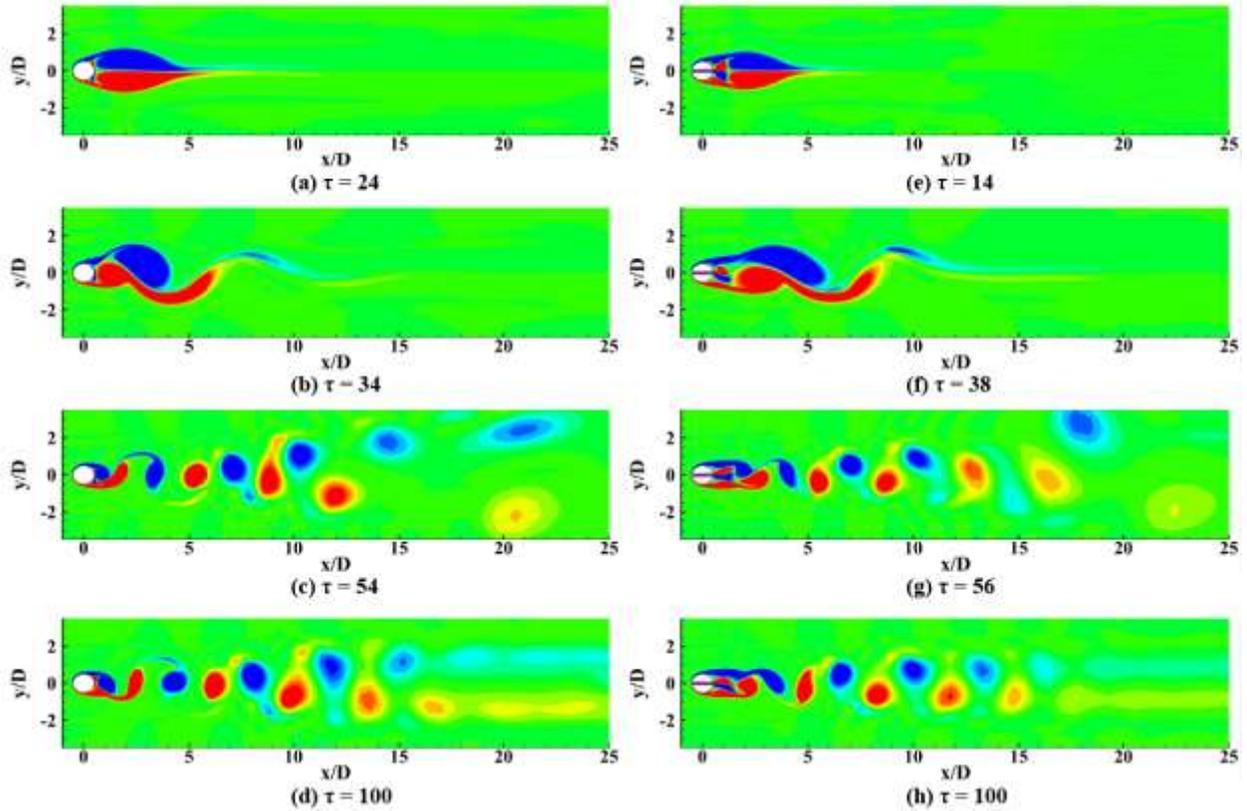

FIG.15. Vorticity contour at time instant (a) $\tau = 24$ (b) $\tau = 34$ (c) $\tau = 54$ (d) $\tau = 100$ for S/D = 0 at Re = 500 and (e) $\tau = 14$ (f) $\tau = 38$ (g) $\tau = 56$ (h) $\tau = 100$ for S/D = 0.15 at Re = 500 (Vorticity legends is kept common for both the case)

TABLE IV. First eight eigenvalues for the demonstration of snapshots convergence

| | Number of snapshots | | |
| --- | --- | --- | --- |
| Mode | 50 | 100 | 150 |
| I | 48.993 | 48.916 | 48.863 |
| II | 47.676 | 47.753 | 47.816 |
| III | 1.543 | 1.538 | 1.534 |
| IV | 1.503 | 1.508 | 1.512 |
| V | 0.1336 | 0.1323 | 0.1313 |
| VI | 0.1289 | 0.1302 | 0.1312 |

## G. Proper orthogonal decomposition (POD)

The proper orthogonal decomposition (POD) technique is a useful mathematical tool to evaluate the dominant flow structure (based on the energy contribution). The POD modes based on the respective fluctuation kinetic energy can be calculated through the instantaneous flow field (snapshots), and the extracted POD modes are proportional to their eigenvalues. In most cases, only the first few modes (93-95% cumulative value of energy in modes) are required to represent the features of flow in the system.

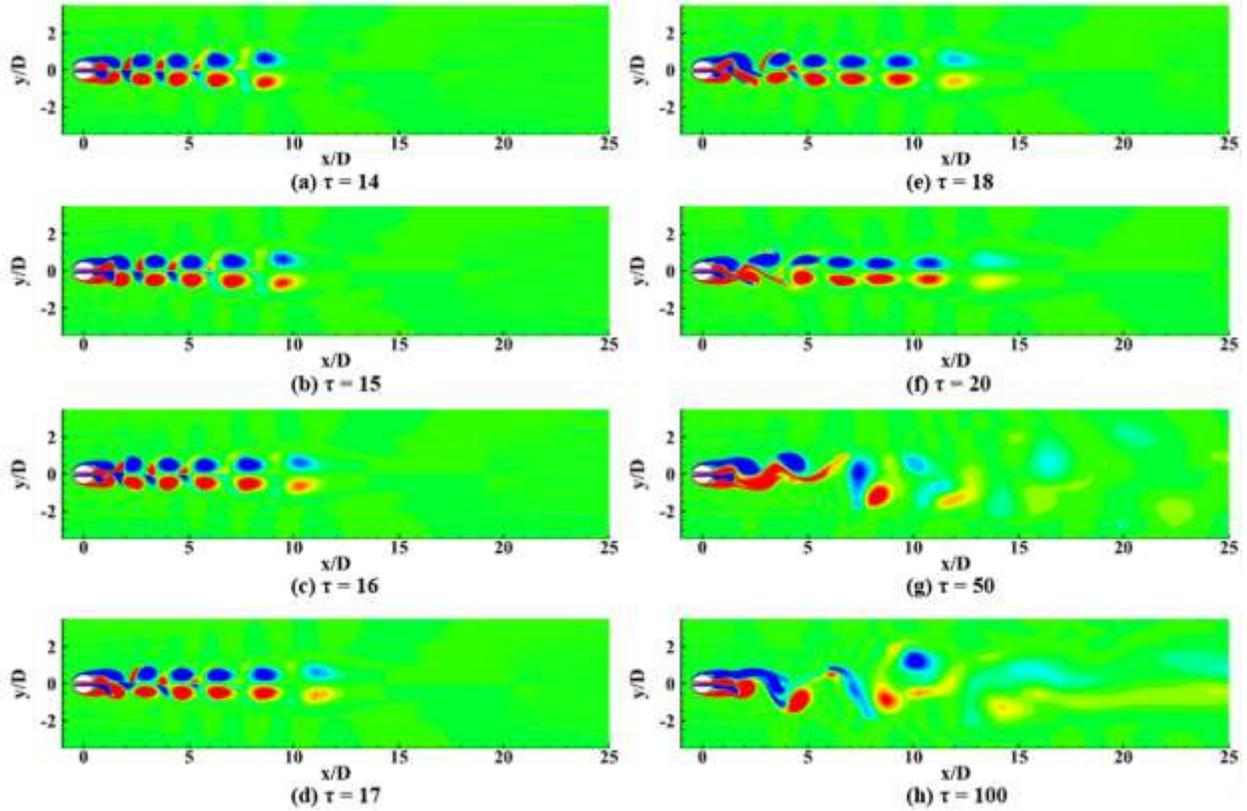

FIG. 16. Vorticity contour at time instant (a) τ = 14 (b) τ = 15 (c) τ = 16 (d) τ = 17 (e) τ = 18 (f) τ = 20 (g) τ = 50 and (h) τ = 100 for S/D = 0.25 at Re = 500 (Vorticity legends is kept common for the case)

The snapshot independence study is performed using the same computational grid (grid-3 in section II-B) to investigate the adequate number of snapshots for POD calculation. Table IV presents the normalized energy of the first six modes from 50, 100, and 150 snapshots POD analysis for S/D=0.25 at Re=100. The first energy mode depicts minor over-prediction by 50 snapshots compared to 100 and 150 snapshots. There is no significant difference in the values of energy modes of 100 and 150 snapshots. The structures of the POD modes show no difference for 50, 100, and 150 snapshots. Hence, 100 snapshots are sufficient enough to carry out the POD analysis for the present study.

Figures 17 (a-d) present the longitudinal velocity over the modified cylinder for the first four POD modes. The first two POD modes contain more than 96% of the total kinetic energy, and these modes express significant flow physics (Table IV). It is noteworthy that these first two modes exhibit similar energy. The structure in modes 1 and 2 shows anti-symmetry (with the equal value and opposite sign about the centerline downstream of the cylinder), evidence of the alternate periodic vortex shedding phenomenon. The POD modes 3 and 4 have a similar structure, but the vortical structures accommodate less energy than modes 1 and 2. Modes 3 and 4 demonstrate the symmetric structure about the centerline downstream of the cylinder, representing the contribution of the shear layer on the flow. Bukka et al.[43] also reports a similar conclusion of pod modes in the study of suppression of vortex-induced vibration (VIV) over the cylinder with a passive flow control device.

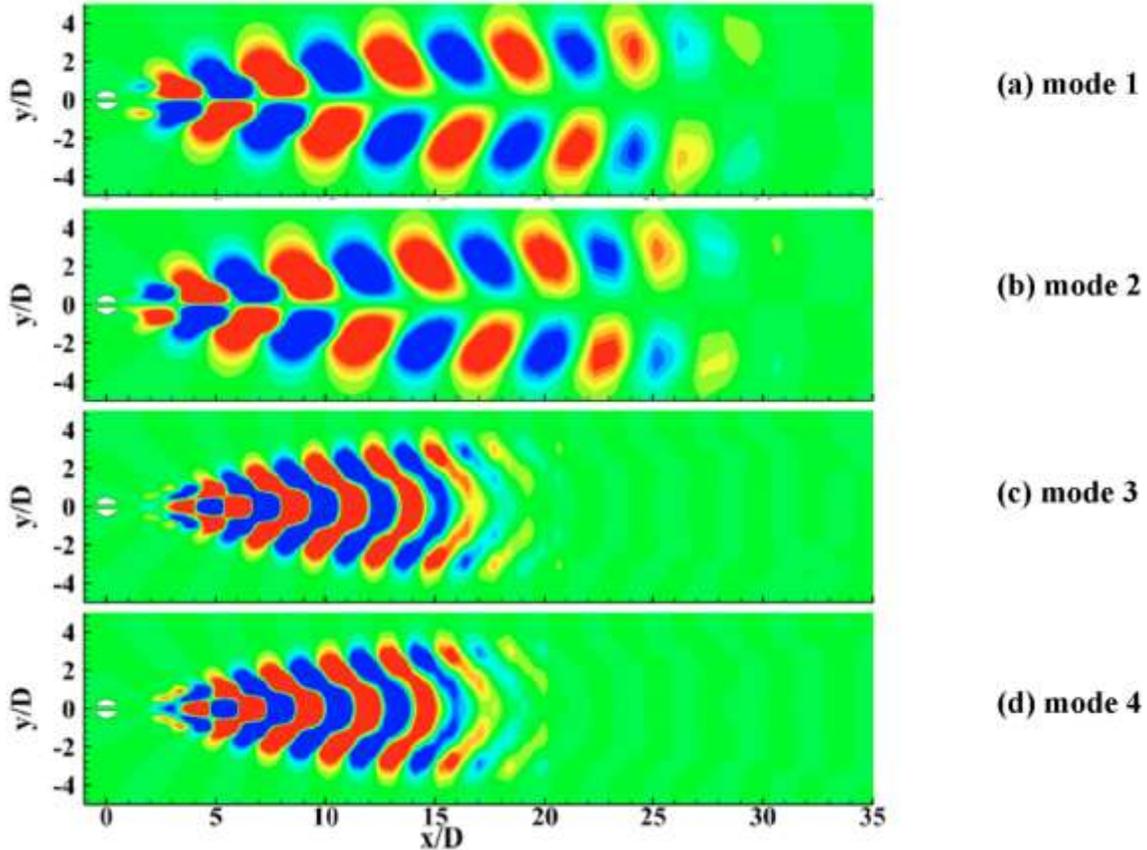

FIG. 17. For the S/D = 0.25 at Re = 100 cases: (a-d) First four most energetic modes for energy-based decomposition; the contour represents the component corresponding vorticity.

The value of the Strouhal number (calculated using the temporal coefficients of POD) is found to be at 0.1641 and 0.1533 for the normal cylinder and S/D = 0.25 case at Re =100, respectively. There is a very minimal difference in the Strouhal number values of the temporal coefficient of POD as compare with the Strouhal number of the lift coefficient, as shown in Section III-A. The same frequency for the temporal coefficient and lift coefficient can also verify that the POD analysis accurately captures the dominant vortical structure of unsteady data.

Further, comparing various slit cylinder cases at Re=500 allows us to have more insight into the flow at this range of Reynolds numbers. The flow shows irregular shedding for S/D> 0.15 at Re=500, and this is the main reason behind selecting the flow investigation on Re = 500. Figure 18 presents the comparison of cumulative energies for the cases (S/D = 0, 0.15, 0.25) at Re = 500. The first mode of the normal cylinder contains the highest energy as compared to the slit cylinder cases. The cumulative energy plot shows fascinating behavior for case S/D = 0.25. The modes contain very distributed energy for the case S/D = 0.25, and It takes 28 modes to complete 93% of total energy. In comparison, only 2 modes contain more than 93% energy for the normal cylinder and S/D=0.15 cases.

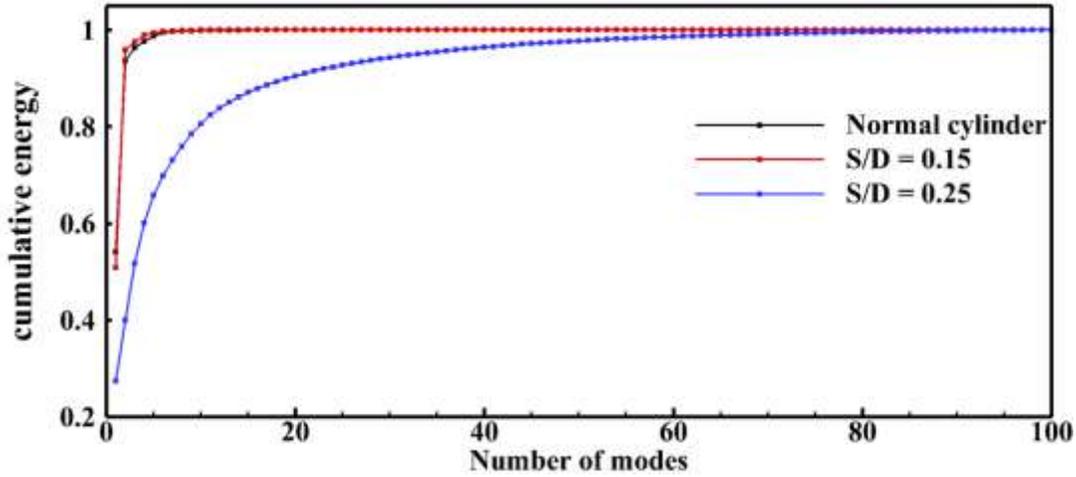

FIG. 18. Cumulative energy of modes for the normal cylinder, S/D = 0.15 and S/D=0.25 at Re =500

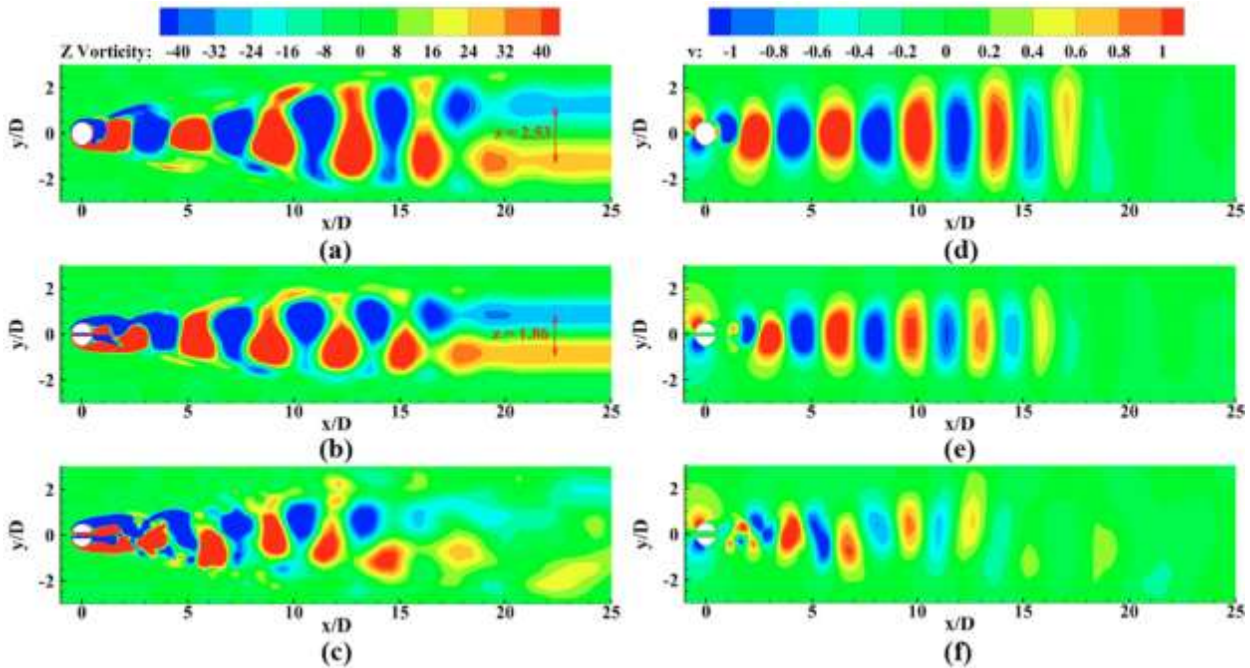

FIG. 19. Reconstructed vorticity structure at Re = 500: (a) Mode 1 and 2 combined for S/D = 0 (b) Mode 1 and 2 combined for S/D=0.15 (c) Mode 1- 28 combined for the S/D = 0.25, Reconstructed transverse velocity (d) Mode 1 and 2 combined for S/D = 0 (e) Mode 1 and 2 combined for S/D=0.15 (f) Mode 1- 28 combined for the S/D = 0.25.

The primary vortex through the cylinder contains most of the energy in the normal cylinder and S/D =0.15 cases. But for S/D=0.25, due to interaction between the primary vortex (through the cylinder) and secondary vortex (through the slit), the energy of the primary vortex is reduced and distributed in irregular shedding. Furthermore, we investigate flow features with the reconstructed vorticity structure modes and reconstructed transverse velocity modes of three cases (S/D = 0, 0.15, and 0.25), as illustrated in Figure 19. The cases (S/D = 0 and 0.15) represent the periodic shedding while S/D = 0.25 exhibits the complex and irregular shedding. The suppression in the case S/D =0.15 can be represented by the transverse spacing of the vortex shedding, which is 1.86D for S/D =0.15 and 2.53D S/D = 0 at Re = 500 (Figures 19-a and b). While applying slit on the normal cylinder (S/D<0.15), the reduction in the transverse spacing is evidence of the vortex suppression for the modified cylinder. The flow for S/D = 0.25 (Figure 19-c) shows non-periodic vortex shedding, so no precise transverse spacing is found in this case.  The irregular and complex flow behavior can be seen from the reconstructed vorticity contour and the transverse velocity contour for S/D =0.25 (Figures 19-c and f). The Strouhal number for the cases S/D = 0, 0.15, 0.25 at Re = 500 are found to be at 0.222, 0.2461, 0.075, which have minor differences with the Strouhal number from the FFT plot of lift coefficient (see Figure 6). The dominant structure of the flow over the modified cylinder (S/D=0.25) is not clear from the POD, so Dynamic mode decomposition (DMD) is applied to investigate the dominant structure.

### H. Dynamic mode decomposition (DMD)

The investigation of DMD provides the coherent structures for normal cylinder and slit cylinder cases at Re =500 concerning dominant frequency. It is more appropriate for a vortical structure investigation to use vorticity (enstropy) for DMD, so the present study also utilizes vorticity-based DMD.  Figure 20 presents the $L_2$-norm plots of eigen function versus frequency for three different cases (S/D=0, 0.15 and 0.25) at Re = 500. One can note the dominant modes and their respective frequencies in the $L_2$-norm plot.

The frequency corresponding to the mean mode is at 0 Hz, which is evident because the mean mode has no fluctuations. The fundamental frequencies of all three cases are approximately equal to the fundamental frequency evaluated by the POD and the FFT of the lift coefficients. For S/D=0 and 0.15 cases, the flow exhibits periodic shedding, which can be confirmed through the $L_2$-norm plots (Figure 20-a and b). The fundamental frequency $L_2$-norm plot displays two more frequencies (first and second harmonics of the flow).  For S/D = 0.25 case, the $L_2$-norm plot shows only one frequency, and no harmonics are found for this case. It can be concluded here that the flow exhibits non-periodic shedding.

Figure 21 presents the real part of dynamic modes corresponding to the fundamental frequency for all three cases. The dominant vortical structure for the case S/D = 0 has the highest vorticity compared to the rest of the slit cylinder cases, and the major vortical structure is found just downstream of the normal cylinder. For the case, S/D = 0.15, the flow coming through the slit pushes the primary vortices downstream, and one can note the small vortical structure through the slot through the cylinder. The flow injection through the slit reduces the strength of the primary vortices, which is the reason for the suppression in the vortex shedding for this case.

The dominant mode for the case S/D=0.25 reveals exciting features of the flow. In S/D = 0.15, the primary vortex possesses strength more than the secondary vortex (vortex through the slit). But for S/D = 0.25, the secondary vortex shows more strength than the primary vortex, which can also be seen in the fundamental frequency difference in both cases (the fundamental frequency for S/D=0.15 is 0.2435 while it is 0.073 for the case S/D=0.25).  The interaction between primary and secondary vortices of the slit cylinder creates irregular and complex shedding downstream of the modified cylinder.

Overall, the mean and instantaneous flow analysis strongly advocates the suppression of vortex shedding by applying slit through the cylinder. The energy-based POD analysis of instantaneous snapshots investigates the dominant vortical structure of the flow over a cylinder, and the results reveal that the dominant vortical structure has more strength in the case of a normal cylinder. The enstrophy-based DMD analysis helps examine the dominant vortical structures of fundamental frequencies and their harmonics, which reports that the suppression in the strength of the vortical structure for a modified cylinder.

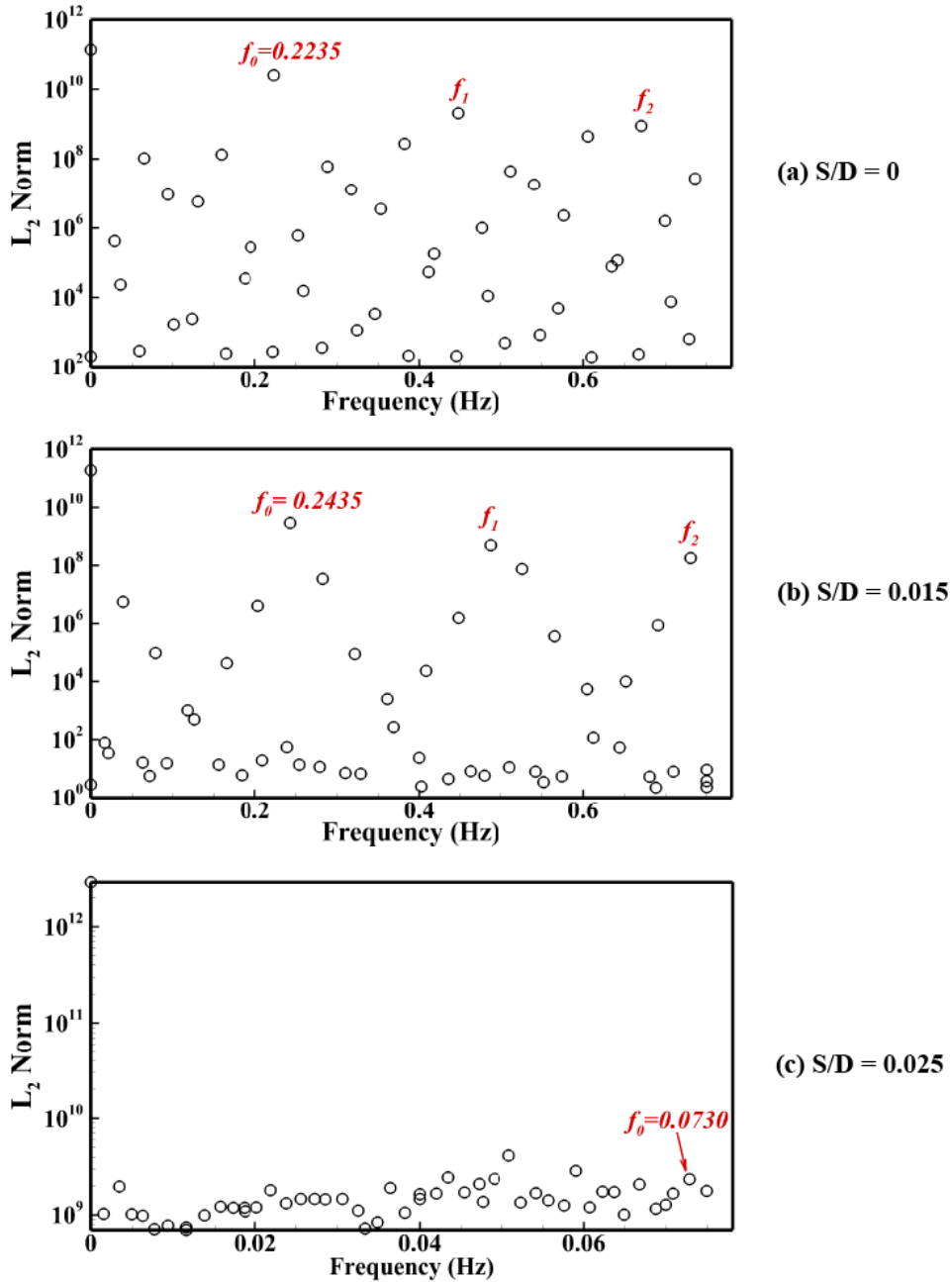

FIG. 20. L2 norm of eigenfunction against frequency corresponding to dominant modes for (a) S/D = 0 (b) S/D = 0.15 (c) S/D = 0.25 at Re = 500.

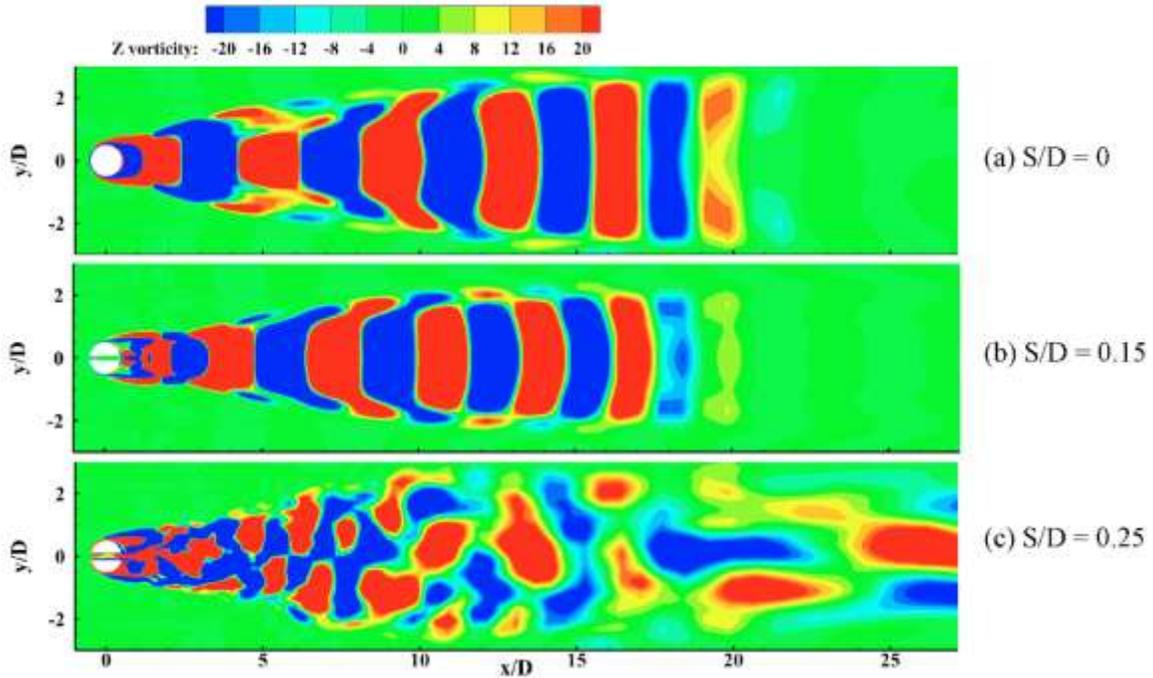
FIG. 21. The real part of dynamic mode corresponding to the fundamental harmonics for (a) S/D = 0 (b) S/D = 0.15 (c) S/D = 0.25 at Re =500

## IV. Conclusion

The present study investigates the suppression of vortex shedding using the passive flow control as a slit through the cylinder for varying slit ratios (S/D = 0 to 0.25) over a wide range of laminar regimes. The slit angle (θ) also varies from 0 to 90º for a few selected S/D (0.05, 0.15, and 0.25) at Re = 100 and 500 to investigate the effect of the slit angle (θ) on the vortex shedding. To validate the simulated data, the predicted results for flow over a normal cylinder at Re = 100 demonstrate good prediction compared to the published literature. The present study also reports flow over the normal and modified cylinder for Reynolds numbers 100 to 500. The extra flow through the cylinder slot increases the coefficient of pressure and delays the shear layer's interaction; thus, the vortex shedding gets suppressed. The global instability downstream of the cylinder reduces with the application of slit on the normal cylinder, and the global instability exhibits a decreasing trend S/D variation. The modified cylinder for all S/D variation shows the periodic vortex shedding behavior similar to the normal cylinder up to Re = 200. The interaction of the primary vortex (cylinder vortex) and secondary vortex (through the slit) starts between Re =200 and 300 for S/D>0.15 and shows periodic vortex shedding behavior. This periodic behaviour maintains up to Re = 320 until the flow demonstrates complex irregular vortex shedding for Re = 330 for S/D =0.20. The flow with a slit width ratio (S/D) less than 0.15 remains periodic, similar to a normal cylinder. The suppression in vortex shedding always happens for the whole laminar regime (Re=100 to 500). For S/D>0.15, flow shows the periodic behavior up to Re=300, which becomes complex irregular shedding for Re>300. The strong interaction of the primary vortex (cylinder vortex) and secondary vortex (through the slit) causes complex and irregular vortex shedding. The variation of the slit angle (θ) shows the monotonous increment in the rms value of the lift coefficient. The behavior of the shedding remains periodic for all the cases except S/D=0.25 at Re = 500. The S/D=0.25 with Slit angle (θ=20º) illustrates the maximum suppression in vortex shedding; henceforth, the rms value of lift coefficient increases with the slit angle. The flow investigation also shows the symmetry-breaking bifurcation point, where the flow behavior changes from symmetric to asymmetric solution at Re=232, and again flow exhibits symmetric solution at Re=304. The proper orthogonal decomposition (POD) analysis on the unsteady data reveals the suppression of vortex shedding by applying a slit through the cylinder as passive flow control. The cumulative energy curve at Re = 500 displays the suppression in vortex shedding for the modified case compared with the normal cylinder. For S/D = 0.25 at Re = 500, the energy of modes is very distributed, and it takes the first 28 modes to cumulated 93% energy while only the first two modes contain more than 93% for other cases. The dynamic mode decomposition (DMD) presents the dominant modes for the fundamental frequency. The dominant vortical structure for the cases (S/D = 0.15 and 0.25) at Re = 500 shows that the primary vortex contains more strength than the secondary vortex for S/D = 0.15, and the secondary vortex incorporates more strength as compared to the primary vortex for the case of S/D = 0.25.

**Data Availability**

The data that support the findings of this study are available from the corresponding author upon reasonable request.

**ACKNOWLEDGMENTS**

Simulations are carried out on the computers provided by the Indian Institute of Technology Kanpur (IITK) (www.iitk.ac.in/cc). The manuscript preparation and data analysis have been carried out using the resources available at IITK. This support is gratefully acknowledged.